\documentclass[twocolumn,showpacs,preprintnumbers,amsmath,amssymb]{revtex4}

\usepackage{graphicx}
\usepackage{dcolumn}
\usepackage{bm}

\begin{document}

\title{
Quantum phase transition in an array of coupled dissipative cavities
}

\author{
Ke Liu,$^{1}$
Lei Tan,$^{1,}$
C.-H. Lv,$^{1}$
and W. M. Liu$^{2}$
}

\affiliation{
$^{1}$Institute of Theoretical Physics, Lanzhou University, Lanzhou 730000, China\\
$^{2}$Beijing National Laboratory for Condensed Matter Physics, Institute of Physics, Chinese Academy of Sciences, Beijing 100190, China
}
\date{\today}

\begin{abstract}
The features of superfluid-Mott insulator phase transition
in the array of dissipative nonlinear cavities are analyzed.
We show analytically that the coupling to the bath can be reduced to
renormalizing the eigenmodes of atom-cavity system.
This gives rise to a localizing effect and drives the system into mixed states.
For the superfluid state,
a dynamical instability will lead to a sweeping to a localized state of photons.
For the Mott state,
a dissipation-induced
fluctuation will suppress the restoring of long-range phase coherence
driven by interaction.
\end{abstract}


\maketitle

One of the remarkable applications of coupled cavity arrays
is to realize quantum simulators~[\onlinecite{SC1, SC2, SC3, CCA}].
Relying on the controllability of optical systems,
it could be useful to attack some unclear physics and
to explore new phenomenon in quantum many-body systems
~[\onlinecite{Glass, MD, Fermi, JCH_Hur, SCT}].
In particular, over the past years the experimental progresses in
engineering strong interaction of photons and atoms
~[\onlinecite{C_M, C_C, C_S, C_B}]
and in fabricating large-scale arrays of
high-quality cavities~[\onlinecite{array1, array2}]
make this potential application may become a reality in the near future .
However, the quantum optical systems in general couple
to an external environment~[\onlinecite{open1, open2}],
which will bring the system out
of equilibrium and profoundly affect the dynamics of interest~[\onlinecite{N-E1, N-E2}].
New important questions thus arise and need to be clarified, e.g.
under the realistically experimental conditions,
how the dissipation and decoherence would behave
in these open systems.

In this paper, we propose a possible
answer to the above question by investigating
the superfluid-Mott insulator phase transition in the array of
dissipative cavities.
We show that the transition shares part features of the
non-dissipative counterparts.
There are still two quantum many-body states can be recognized
as the delocalized and localized of photons.
However, very differently,
the dissipation and the decoherence
give rise to a localizing effect and drive the system into
mixed states.
For the superfluid state,
a non-equilibrium dynamical instability
can lead to a sweeping to a localized state at
a finite time.
For the Mott state, where photons are already localized at
each lattice site, the localization holds
but a dissipation-induced fluctuation of photon number
acted on each lattice site will
suppress the restoring of long-range phase coherence.

\begin{figure}[b]
\includegraphics[clip]{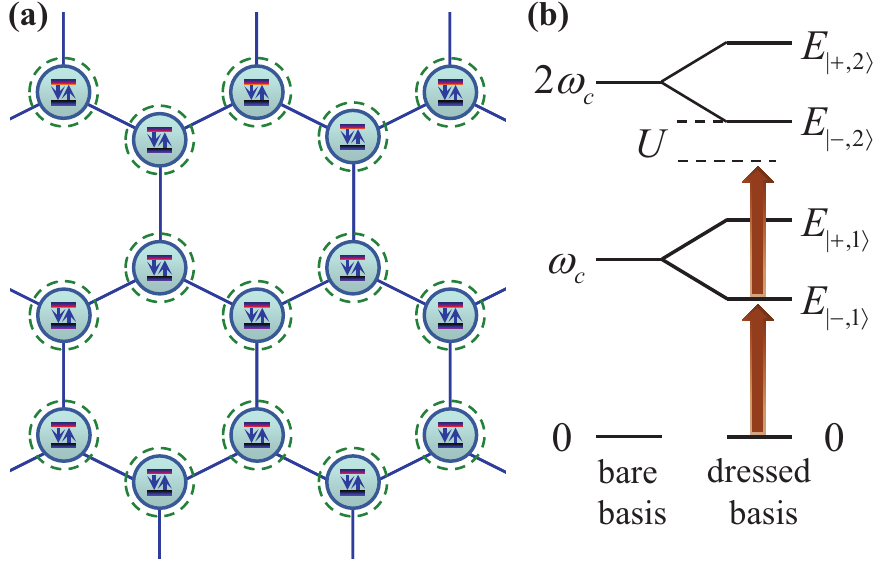}
\caption{
\label{fig:epsart}
A type of possible topologies for two-dimensional cavity arrays,
for $z$ nearest neighbors.
(a) Individual cavities are coupled resonantly to each other due to the
overlap of the evanescent fields. Each cavity contains a two-level system
coupled strongly to the cavity field and
immerses in a bosonic bath(marked by the dash line).
(b) Energy eigenvalues of individual cavity-atom system
on each site. $\omega_{c}=\omega_{a}$ is assumed for simplicity.
The anharmonicity of the Jaynes-Cummings energy levels
can effectively provide an on-site repulsion $U$
to block the absorption for the next photon.
}
\end{figure}
Consider a system consisted of atoms and cavities
coupled weakly to a bosonic environment at zero
temperature.
As the size of individual cavities is generally much
smaller than their spacing,
we assume the photons emitted from each cavity
are uncorrelated. The total Hamiltonian therefore reads
\begin{equation}
H=H_{s}+H_{bath}+H_{coup}.
\end{equation}
where $H_{s}$ is the Hamiltonian for the system,
$H_{bath}=\sum_{j}\sum_{\alpha,k}\omega_{k_{\alpha}}r^{\dag}_{j,k_{\alpha}}r_{j,k_{\alpha}}$
the Hamiltonian for environment, and
$H_{coup}=\sum_{j}\sum_{\alpha,k}(\eta^{\ast}_{k_{\alpha}}r^{\dag}_{j,k_{\alpha}}\alpha_{j}+h.c.)$
the coupled term.
$\alpha=a, c$ labels the operators and physical quantities associated with atoms and
cavities, respectively. $\omega_{k_{\alpha}}$ denotes the frequency of
environmental modes,
$r^{\dag}_{j,k_{\alpha}}$ and $r_{j,k_{\alpha}}$the creation and annihilation operators of
quanta in the $k_{\alpha}$th model
on the $j$th lattice site, and $\eta_{k_{\alpha}}$ the coupling strength.
Here we set $\hbar=1$.

The system we modeled, as depicted in Fig. 1,
is a two-dimensional array of resonant optical cavities,
each embedded with a two-level (artificial) atom coupled strongly to the cavity field.
The possible realizations such as photonic bandgap cavities and
superconducting stripline resonators et al.~[\onlinecite{CCA}].
With $\omega_{a}$ and $\omega_{c}$ being the frequency of atom transition
and cavity mode respectively, in the rotating wave approximation(RWA),
such individual atom-cavity system on site $j$ is well described by
the Jaynes-Cummings Hamiltonian,
$
H^{JC}_{j}=\omega_{c}b^{\dag}_{j}b_{j}+
\omega_{a}\sigma^{+}_{j}\sigma^{-}_{j}+
\beta(\sigma^{+}_{j}b_{j}+h.c.)
$.
Here $b^{\dag}_{j}$ and $b_{j}$
($\sigma^{+}_{j}$, $\sigma^{-}_{j}$)
are photonic(atomic pseudo-spin) rasing
and lowering operators, respectively, $\beta$ the coupled strength.
In the grand canonical ensemble,
$H_{s}$ is therefore given by combing $H^{JC}_{j}$ with photonic hopping term
and chemical potential term,
\begin{equation}
H_{s}=\sum_{j}H^{JC}_{j}-\sum_{\langle j,j'\rangle}\kappa_{jj'}b^{\dag}_{j}b_{j'}-\sum_{j}\mu n_{j}.
\end{equation}
$\kappa_{jj'}$ is the photonic hopping rate between cavities.
Since the evanescent coupling between cavities decreases with the distance exponentially,
we restrict the summation $\sum_{\langle j,j'\rangle}$ running
over the nearest-neighbors.
$
n_{j}=b^{\dag}_{j}b_{j}+\sigma^{+}_{j}\sigma^{-}_{j}
$
counts the total number of atomic and photonic
excitations on site $j$.
$\mu$ is the chemical potential, where the assumption $\mu=\mu_{j}$
for all sites has been made.

Due to the strong coupling, as shown in Fig. 1(b),
the resonant frequencies of individual
atom-cavity system are split into
$
E_{|\pm,n\rangle}=n\omega_{c}\pm\sqrt{n\beta^{2}+\frac{\Delta^{2}}{4}}-\frac{\Delta}{2},
$
where $|\pm,n\rangle$ labels the positive(negative) branch of dressed states,
$\Delta=\omega_{c}-\omega_{a}$ is the detuning.
The anharmonicity of the Jaynes-Cummings energy levels can effectively
provide a on-site repulsion. For instance,
the resonant excitation by a photon with frequency $E_{|\pm,1\rangle}$ will
prevent the absorption of a second photon at $E_{|\pm,1\rangle}$,
which is the striking effect known as photon blockade~[\onlinecite{C_B}].
It is therefore feasible to realize a quantum simulator in terms of
the system described by Eq. (2).
This so called Jaynes-Cummings-Hubbard(JCH) model is recently suggested
by Greentree et al.~[\onlinecite{SC2}].

However, the situation changes dramatically
once taking the degrees of freedom of environment into consideration,
as described by Hamiltonian(1).
A non-equilibrium dynamics for open quantum many-body system do arise,
which is a formidable task to solve.
Here we propose a new method to
eliminate those external degrees of freedom.
To approach this, we regroup Hamiltonian(1) as
\begin{equation}
H=H_{local}-\sum_{\langle j,j'\rangle}\kappa_{jj'}b^{\dag}_{j}b_{j'}-\sum_{j}\mu n_{j},
\end{equation}
where
$H_{local}=\sum_{j}H^{JC}_{j}+H_{bath}+H_{coup}$.

First considering the case that the $j$th cavity contained a initial photon interacts with
a bath, the dynamics is governed by
\begin{equation}
H_{j}=\omega_{c}b^{\dag}_{j}b_{j}+\sum_{k}\omega_{k_{c}}r^{\dag}_{j,k_{c}}r_{j,k_{c}}
+\sum_{k}(\eta^{\ast}_{k_{c}}r^{\dag}_{j,k_{c}}b_{j}+h.c.).
\end{equation}
We denote its eigenvalue as $\omega$ and
expand the eigenvector $|\phi_{j}\rangle$ as
$
|\phi_{j}\rangle=e_{c}b^{\dag}_{j}|\emptyset\rangle
+\sum_{k}e_{k}r^{\dag}_{k_{c}}|\emptyset\rangle
$.
$e_{c}$ and $e_{k}$ are the probability amplitudes for the excitation occupied by
cavity field and environment, respectively. $|\emptyset\rangle$ denotes the vacuum state.
Deducing the equations of these two amplitudes, one can express
$e_{k}$ in terms of $e_{c}$ and integrate out the degrees of freedom of environment
when the coupling to environment is weak,
thus obtain
$
(\omega_{c}+\delta\omega_{c}-i\gamma_{c})e_{c}=\omega e_{c}.
$
$\delta\omega_{c}$ is known as an analog to the Lamb shift in atomic physics
and is sufficiently small.
$\gamma_{c}$ is the decay rate and indicates a finite lifetime
of cavity mode.
We note that the above treatment is precise
under the Born-Markov approximation~[\onlinecite{book}].

This motivates us to introduce a quasi-boson described by $B_{j}$ with
a complex eigenfrequency
$
\Omega_{c}=\omega_{c}-i\gamma_{c}
$,
where $\delta\omega_{c}$ has been absorbed into $\omega_{c}$,
to redescribe the cavity field coupled with a bath
in terms of
$
H^{eff}_{j}|\phi_{j}\rangle=\Omega_{c}|\phi_{j}\rangle
$.
$
H^{eff}_{j}=\Omega_{c}B^{\dag}_{j}B_{j}
$
is the effective Hamiltonian
and now
$
|\phi_{j}\rangle=e_{c}B^{\dag}_{j}|\emptyset\rangle
$
denotes the time-dependent damped basis~[\onlinecite{D-Basis}].
Because of loss, the system would be nonconservative and corresponding operators
would be non-Hermitian. The commutation relation of $B_{j}$ reads
$[B_{j},B^{\dag}_{j'}]=(1+i\frac{\gamma_{c}}{\omega_{c}})\delta_{jj'}$.
Recognizing $\frac{\gamma_{c}}{\omega_{c}}$ is in order of $\frac{1}{Q}$,
with $Q$ being the quality factor of individual cavity.
The bosonic commutation relation is therefore approximately satisfied for
the high-$Q$ cavity, which can be met in most experiments about
cavity quantum electrodynamics(QED).

The complex eigenfrequency underlines the facts that, on one hand,
dissipation is the inherent property for realistic cavity.
When a photon with certain frequency has been injected into a dissipative cavity,
the composite system can not be characterized only by the mode of
cavity field, however, we must take the impacts of environment
into account. On the other hand, in general we do not concern the
time evolution of bath. In this way, the array of dissipative cavities
can be regarded as a configuration consisted of quasi-bosons.
Quite similar operations can be performed on atom to introduce another
kind of quasinormal mode described by
$\tilde{\sigma}^{\pm}_{j}$ with the frequency
$\Omega_{a}=\omega_{a}-i\gamma_{a}$,
where $\gamma_{a}$ is the atomic decay rate.

We can therefore rephrase Hamiltonian (3) with the
renormalized terms,
\begin{equation}
H=\sum_{j}H^{eff}_{j}-\sum_{<j,j'>}\kappa_{jj'}B^{\dag}_{j}B_{j'}-\sum_{j}\mu n_{j},
\end{equation}
with now
$
H^{eff}_{j}=\Omega_{c}B^{\dag}_{j}B_{j}+
\Omega_{a}\tilde{\sigma}^{+}_{j}\tilde{\sigma}^{-}_{j}+
\beta(\tilde{\sigma}^{+}_{j}B_{j}+h.c.)
$
and
$
n_{j}=B^{\dag}_{j}B_{j}+\tilde{\sigma}^{+}_{j}\tilde{\sigma}^{-}_{j}
$.
One nice feature of Hamiltonian (5) is now the losses
describe by leaky rates $\gamma_{a}$ and $\gamma_{c}$
but not by operators. Without having to mention the external degrees of freedom,
this effective treatment would be of great conceptual and, moreover,
computational advantage rather than the general treatment as Hamiltonian (1).
A more microscopic consideration points out that,
in cavity QED region, since the atom is dressed by cavity field,
the atom and field act as a whole subject to a total decay rate $\Gamma$~[\onlinecite{decay}].
In particular,
$\Gamma=n(\gamma_{a}+\gamma_{c})$
for
$\Delta=0$.

To gain insight over the role of dissipation in the superfluid-Mott insulator phase transition,
we use a mean field approximation which could give reliable results
comparing to the Monte Carlo calculations
if the system is at least two-dimensional~[\onlinecite{MF}].
We introduce a superfluid parameter,
$\psi=Re\langle B_{j}\rangle=Re\langle B^{\dag}_{j}\rangle$.
In the present case, the expected value of $B_{j}(B^{\dag}_{j})$
is in general complex with the formation
$
\langle B_{j}\rangle=\psi-i\psi_{\gamma}
(\langle B^{\dag}_{j}\rangle=\psi+i\psi_{\gamma})
$.
$\psi_{\gamma}$ is a solvable small quantity as a function of
$\gamma_{a}$ and $\gamma_{c}$, and vanishes in the limit of no loss.
Using the decoupling approximation,
$
B^{\dag}_{j}B_{j'}=\langle B^{\dag}_{j}\rangle B_{j'}
+\langle B_{j'}\rangle B^{\dag}_{j}
-\langle B^{\dag}_{j}\rangle\langle B^{\dag}_{j'}\rangle
$,
the resulting mean-field Hamiltonian can be written as a sum over single sites,
\begin{equation}
H^{MF}=\sum_{j}
\{
H^{eff}_{j}-z\kappa\psi(B^{\dag}_{j}+B_{j})+z\kappa|\psi|^{2}-\mu n_{j}+O(\psi^{2}_{\gamma})
\},
\end{equation}
where we have set the intercavity hopping rate $\kappa_{jj'}=\kappa$
for all nearest-neighbors with $z$ labeling the number.
For zero temperature, this mean-field approximation is equivalent
to the Gutzwiller approximation, which assumes the wave function of system as
a product of single-site wave function~[\onlinecite{GW}].

$\psi$ can be examined analytically in terms of the second-order perturbation theory,
with respect to the damped dressed basis.
For energetically favorable we assume each site is prepared in
the negative branch of dressed state.
But because the dressed basis is defined on $n\geq1$,
a ground state $|0\rangle$ with the energy $E_{|0\rangle}=0$ need to be supplemented.
Thus
\begin{equation}
\psi=e^{-\Gamma t}\sqrt{-\frac{\chi}{z\kappa\Theta}}.
\end{equation}
$\chi$ and $\Theta$ are functions of all the parameters of the whole system.
Since the evanescent parameter $\kappa$ is a typical small quantity
in systems of coupled cavities,
the perturbation theory gives good qualitative
and even quantitative descriptions
comparing to the numerically results
given by explicitly diagonalizing~[\onlinecite{SC2,SF-MI}].

Arguably the most interesting situation is the
effective photon-photon interactions are
maximized, namely, cavities on resonant with atoms and
with one initial excitations per lattice site~[\onlinecite{JCH}].
And thereby
$
\Gamma=\gamma_{a}+\gamma_{c}=\gamma
$.
With
$
F_{1}=\omega_{c}-\beta-\mu
$
and
$
F_{2}=-\omega_{c}+(\sqrt{2}-1)\beta+\mu
$,
in eq. (7),
$
\Theta=\frac{1}{2F^{2}_{1}+2\gamma^{2}}
+\frac{3+2\sqrt{2}}{4F^{2}_{2}+4\gamma^{2}}>0
$,
and
\begin{equation}
\chi=\frac{F_{1}}{2F^{2}_{1}+2\gamma^{2}}
+\frac{(3+2\sqrt{2})F_{2}}{4F^{2}_{2}+4\gamma^{2}}
+\frac{1}{z\kappa e^{-2\gamma t}}.
\end{equation}
In the absence of loss, one can recognize $\chi=0$ is the well known
self-consistent equation and therefore distinguish the superfluid phase and Mott phase.
Nevertheless, the coupling to environment inducing a non-equilibrium dynamics,
thus no strict phase exists.
However, provided the external time dependence is much slower than
the internal frequencies of system,
there remains two fundamentally different quantum state can be
identified through whether $\psi$ vanishes or has a finite value,
i.e. photons localized in each lattice site
and delocalized across the cavities.

\begin{figure}
\includegraphics[clip]{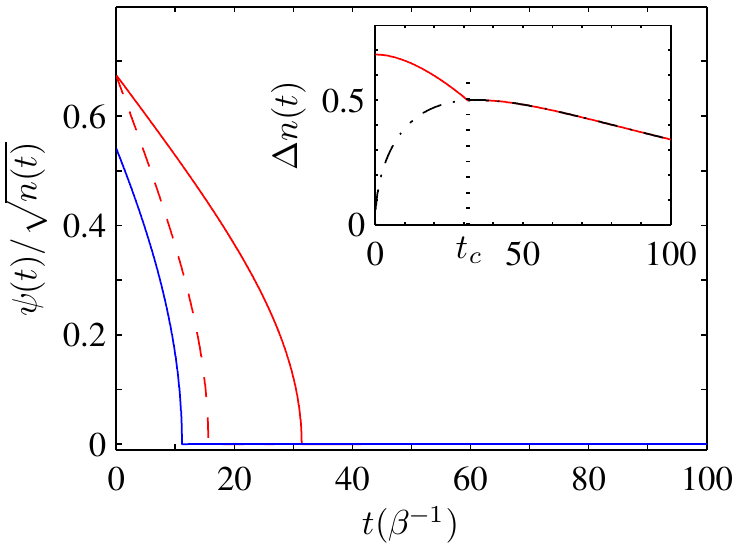}
\caption{
\label{fig:epsart}
The time dependence of the long-range phase coherence
and the photon number fluctuation on each site for a certain
initial state(inset).
For a initial superfluid state
(the red line for $\frac{z\kappa}{\beta}=0.3$
and blue line for $\frac{z\kappa}{\beta}=0.2$),
before $t_{c}$ the long-range order decays continuously
and the fluctuation is a total effect of photon hopping
and photon leakage
(the solid line for $\frac{\gamma}{\beta}=0.01$
and dash line for $\frac{\gamma}{\beta}=0.02$).
Beyond $t_{c}$, the superfluidity breaks down and
the fluctuation behaves as the localized state
(the dot-dash line for $\frac{z\kappa}{\beta}=0$
and $\frac{\gamma}{\beta}=0.01$).
}
\end{figure}
To analyze the physics of the transition between these two
states in detail, we proceed our discussion from two aspects.
First, we start with the superfluid phase
and track the time evolution of long-range phase coherence.
The prefactor $e^{-\Gamma t}$ in Eq. (7) indicates
the expected decay of $\psi$.
However, more importantly,
a dynamical instability due to the coupling to the external environment
is revealed by $\chi$.
As illustrated in Fig. 2,
for $t\ll\beta^{-1}$,
$\psi$ has a slightly reduction scaled by $\frac{\gamma^{2}}{\beta^{2}}$.
For $t>\beta^{-1}$,
$z\kappa e^{-2\gamma t}$ is the leading term
and pronounces the decrease of effective tunneling energy.
Consequently, a photon hopping rate $\kappa$ given initially
in the superfluid region will cross the critical point
at a time
$
t_{c}\simeq\frac{1}{2\gamma}\ln\frac{\kappa}{\kappa_{c}}
$,
with $\kappa_{c}$ being the critical tunneling energy
for a given $z$ and $\beta$.
Before $t_{c}$, a non-local region is still recognized as non-local.
The dissipation has not changed the fundamental nature of the system,
albeit with the reduction of long-range phase coherence
and an additional fluctuation due to photon leakage.
Nevertheless, beyond $t_{c}$, the superfluidity breaks down,
i.e. a transition to the localized state do occur.
An analogous localizing effect is described in an
optical lattice system very recently,
where the spontaneous emission of atoms owing to
the lattice heat leads to decoherence of
many-body state~[\onlinecite{localization}].

In what follows, in contrast, we start in the Mott state
and discuss the impacts of dissipation on
the critical behavior and the fluctuation behavior.
Consider the initial state is deep in the Mott phase,
$
\frac{z\kappa}{\beta}=0
$,
and we continuously increase the intercavity coupled rate.
For the related ideal case, one can reach the superfluid phase at
$
\frac{z\kappa}{\beta}=(\frac{z\kappa}{\beta})'_{c}\simeq0.16
$.
However, the presence of bath converts coherences originally
in the system into entanglement of the system and
the environment~[\onlinecite{C_C}],
thus the effective tunneling energy will be lower than expected.
Moreover, this impact will accumulate along with time.
As shown in Fig. 3,
to expect the appearance of photonic hopping
we must keep increasing $\kappa$.
On the other hand,
despite the long-range order is still absent,
different from the pure Mott state,
there will be a fluctuation owing to photon leakage
acted on each lattice site(dot-dash line in Fig. 2).
Consequently, we will not be able to restore
the long-range phase coherence perfectly
by driven $\frac{z\kappa}{\beta}$ into the superfluid region.

\begin{figure}
\includegraphics[clip]{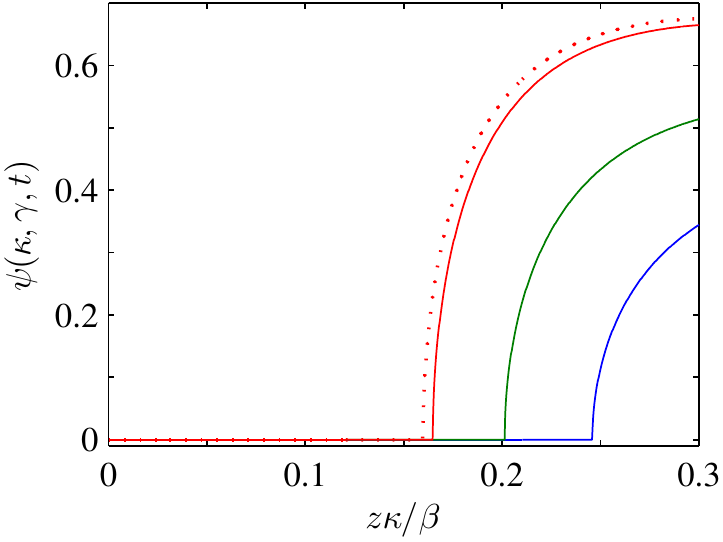}
\caption
{
\label{fig:epsart}
The restoring of long-range phase coherence
from the Mott state.
Influences of dissipation
depend on the leaky rate $\gamma$
(the dot and solid line for $\frac{\gamma}{\beta}=0$ and $0.05$, respectively)
and will accumulate along with time
(the red, green, and blue lines for $t=0$, $0.1\gamma^{-1}$,
and $0.2\gamma^{-1}$, respectively).
}
\end{figure}

In summary, we have shown analytically the features of
superfluid-Mott insulator phase transition
in the array of dissipative cavities.
Our analysis sufficiently takes
into account the intrinsically dissipative nature of
open quantum many-body system, and identifies how dissipation
and decoherence would come into play.
For the further experimental signature,
we predict that there will be a localizing effect.


\end{document}